%% file: Blnugamma.tex
\RequirePackage{lineno}
\documentclass[aps,twocolumn,superscriptaddress,showpacs,preprintnumbers,amsmath,reprintamssymb,nofootinbib,a4paper]{revtex4-1}
\usepackage{scrextend}
\usepackage{graphicx} 
\usepackage{dcolumn}  
\usepackage{subcaption}
\usepackage[font=small, labelfont=bf, justification=raggedright, format=plain]{caption}

\graphicspath{{ps}}

\usepackage{header3}
\usepackage[italic]{hepnames}
\usepackage{booktabs}     
\usepackage{hhline}
\usepackage{multirow}

\input{shorts}
\newcounter{FBC}

\newcounter{MGC}

\newcounter{PGC}

\begin{document}


\preprint{\vbox{ \hbox{   }
						\hbox{\bf Belle Preprint {\bf 2018-24}}
						\hbox{\bf KEK Preprint {\bf 2018-67}}
}}

\title{ \quad\\[1.0cm] Search for the rare decay of \boldmath{\btolng} with improved hadronic tagging}

\input{authors}

\begin{abstract}

We present the result of the search for the rare \PB meson decay of \btolng with $\ell =\Pe,\mu$. For the search the full data set recorded by the Belle experiment of $\SI{711}{{fb}^{-1}}$ integrated luminosity near the $\PUpsilonFourS$ resonance is used. Signal candidates are reconstructed for photon energies \Eg larger than $\SI{1}{GeV}$ using a novel multivariate tagging algorithm. The novel algorithm fully reconstructs the second \PB meson produced in the collision using hadronic modes and was specifically trained to recognize the signal signature in combination with hadronic tag-side \PB meson decays. This approach greatly enhances the performance. Background processes that can mimic this signature, mainly charmless semileptonic decays and continuum processes, are suppressed using multivariate methods. The number of signal candidates is determined by analyzing the missing mass squared distribution as inferred from the  signal side particles and the kinematic properties of the tag-side \PB meson. No significant excess over the background-only hypothesis is observed and upper limits on the partial branching fraction $ \Delta \mathcal{B} $ with \EgOneGev individually for electron and muon final states as well as for the average branching fraction of both lepton final states are reported. We find a Bayesian upper limit of \BFuplep at 90\% CL and also report an upper limit on the first inverse moment of the light-cone distribution amplitude of the $\PB$ meson of \lambdab at 90\% CL.



\end{abstract}

\pacs{13.20.He, 14.40.Nd}

\maketitle

\tighten


\section{Introduction}
The precision study of leptonic \PB meson decays offers a unique path to test the Standard Model (SM) of particle physics: new heavy mediators or sterile neutrinos could contribute to the decay amplitudes and lead, for instance, to lepton flavor universality breaking effects. In the SM, however, leptonic transitions are helicity suppressed, making an observation in final states involving electrons or muons challenging. The helicity suppression can be lifted, if one considers final states which involve an additional photon in the final state, emitted for instance from the light $\Pup$ quark. The decay rate for such \btolng processes \cite{CC} is suppressed by a factor of $\alpha_{\rm em}$ and the resulting decay amplitude \cite{NU}  may be expressed as
\begin{align}
 \mathcal{A}(\btolng) = & \frac{G_\mathrm{F} V_{ub}}{4\sqrt{2}} \langle \bar \ell \nu \gamma | \ell \gamma^\mu P_\mathrm{L} \bar \nu \cdot \Pup \gamma_\mu P_\mathrm{L} \APbottom | \APB \rangle  \, ,
\end{align}
with the projection operator $P_\mathrm{L} = (1 - \gamma_5) / 2$, $G_\mathrm{F}$ denoting Fermi's constant, $\ell$ denoting either an electron or a muon field, $\Pup$ and $\APbottom$ as quark fields and \Vub denoting the relevant Cabibbo-Kobayashi-Maskawa (CKM) \citep{Kobayashi:1973fv, Cabibbo:1963yz} matrix element of the transition. The hadronic transition can be fully described by two form factors parameterizing the axial-vector and vector hadronic currents, denoted by $F_\mathrm{A}$ and $F_\mathrm{V}$, respectively. The differential decay rate as a function of these two form factors and the photon energy \Eg is given by 
\begin{align}\label{eq:rate}
 \frac{\text{d} \Gamma(\btolng)}{\text{d} \Eg} = & 
\frac{\alpha_{\rm em}G_\mathrm{F}^2|V_{ub}|^2}{6\pi^2}m_{\PB} \Eg^3
\left(1-\frac{2\Eg}{m_{\PB}}\right) \nn \\
& \, \times \left(\,\Big|F_\mathrm{V}\Big|^2+\Big|F_\mathrm{A} + \frac{e_\ell f_{\PB}}{\Eg}\Big|^2\,
\right)\,,
\end{align}
with $e_\ell$ denoting the charge of the lepton and $ m_{\PB} $ the \PB meson mass. At high photon energies of \EgOneGev, both form factors can be expanded \citep{Beneke:2011nf} as
\begin{align}
F_\mathrm{V}(\Eg) &= 
\frac{e_{\Pup} f_{\PB} m_{\PB}}{2 \Eg \lb(\mu)} R(\Eg,\mu) +  \xi(\Eg)  +  \Delta\xi(\Eg)\,, \\
\nonumber\\
F_\mathrm{A}(\Eg) &= 
\frac{e_{\Pup} f_{\PB} m_{\PB}}{2 \Eg \lb(\mu)} R(\Eg,\mu)  +  \xi(\Eg) - \Delta\xi(\Eg)\,, 
\label{eq:FFs}
\end{align}
with $f_{\PB}$ denoting the \PB meson decay constant and $e_{\Pup}$ the charge of the $u$ quark line. The factor $R(\Eg, \mu)$ accounts for photon emissions from the light spectator quark in the \PB meson and is unity at tree level. The $\xi$ and $\Delta \xi$ terms are power suppressed by $1/m_{\Pbottom}$ and $1/(2 \Eg )$ and contain a symmetry conserving and breaking part for both form factors $F_\mathrm{A}$ and $F_\mathrm{V}$. In Refs.~\citep{Wang:2018wfj,Beneke:2018wjp} the leading contributions to $\xi$ and $\Delta \xi$ were evaluated and predictions with an accuracy of $\mathcal{O}(20\%)$ were presented. The parameter \lb is related to the first inverse moment of the leading-twist \PB meson light-cone distribution amplitude $\phi_+$ in the high energy limit, $\lb^{-1} = \int_0^\infty \, \text{d} w \, \phi_+(w)$ with $w$ denoting the light-cone momentum. This parameter is of great relevance and serves as an important input to understand QCD factorization used to predict non-leptonic \PB meson decays~\citep{Grozin:1996pq,Beneke:2000ry,Beneke:2000wa}. The partial branching of \btolng is expected to be of the order of $ \mathcal{O}(10^{-6}) $ for photon energies of \EgOneGev and \lb in the range of several hundred MeV \citep{Beneke:2011nf}.

The search for the rare \btolng decay thus serves two purposes: first a limit or observation of the partial branching fraction can be used in combination with the decay rate Eq.~\ref{eq:rate} and input from theory for $\xi$ and $\Delta \xi$, as well as for $f_{\PB}$ and \Vub, to experimentally determine the value of \lb. Second, with the Belle~II experiment the \btolng decay rate could offer a future additional path to determine \VubAbs by using lattice calculations for \lb. In this manuscript we present a novel method to determine \lb, which uses the measured ratio of \btolng with respect to \btopln. This cancels the explicit dependence of \Vub on \lb and in the ratio several experimental uncertainties cancel. 

The search in this manuscript supersedes the earlier result of Ref.~\citep{Heller:2015vvm}, using an improved hadronic tagging and more accurate modeling of the charmless semileptonic backgrounds. The hadronic tagging is based on the Full Event Interpretation (FEI)~\citep{Keck:2018lcd}, which uses multivariate methods trained to recognize the signal signature (consisting of a single photon with \EgOneGev and a lepton candidate) in conjunction with a hadronic \PB meson decay. This signal-specific approach enhances the selection efficiency in comparison to the previous analysis by a factor of three, which results in an improvement of $1.9 \sigma$ for the expected significance for a given partial branching fraction of \ensuremath{ \Delta \mathcal{B}( \PB^{+} \to \Plepton^{\, +} \Pnulepton \Pgamma) = 5.0 \times 10^{-6}}\xspace and with a similar ratio of signal over background events.  Backgrounds that can mimic signal, most importantly \btopln and other charmless semileptonic decays, are suppressed with a dedicated $\pi^0$ and $\eta$ veto, which combines the signal photon with other calorimeter clusters to form a multivariate classifier. The signal is extracted by analyzing the missing mass squared distribution, \Mmisssq, which for correctly reconstructed signal decays peaks around $0 \, \text{GeV}^2$. Semileptonic background and event candidates from continuum processes are shifted towards positive values. The \btolng branching fraction is extracted simultaneously with a \btopln control sample, which constrains the most dominant background contribution. The extracted yields are converted into (partial) branching fractions to determine the ratio of \btolng to \btopln decays. 

This manuscript is organized as follows: Section~\ref{sec:selection} details the analysis strategy and selection. Section~\ref{sec:stats} summarizes the statistical analysis of the \Mmisssq distribution and the limit setting procedure. In Section~\ref{sec:fei} the calibration procedure of the multivariate tagging algorithm is summarized. Sections~\ref{sec:syst}~and~\ref{sec:results} discuss the systematic uncertainties and the main results. The manuscript concludes with a summary in Section~\ref{sec:summary}.

\vspace{2ex}
\section{Data set and Analysis Strategy}\label{sec:selection}

The full Belle data set of \mbox{$(772 \pm 10) \times 10^6$} \PB meson pairs is analyzed, produced at the KEKB accelerator complex~\citep{KEKB} at a center-of-mass energy of $\SI{10.58}{GeV}$ at the $\PUpsilonFourS$ resonance. The Belle detector is a large-solid-angle magnetic spectrometer that consists of a silicon vertex detector (SVD), a 50-layer central drift chamber (CDC), an array of aerogel threshold  \v{C}erenkov counters (ACC),  a barrel-like arrangement of time-of-flight scintillation counters (TOF), and an electromagnetic calorimeter comprised of CsI(Tl) crystals (ECL) located inside a superconducting solenoid coil that provides a \SI{1.5}{T} magnetic field. An iron flux return located outside of the coil is instrumented to detect \PKlong mesons and to identify muons (KLM). A more detailed description of the detector can be found in Ref.~\citep{Belle}.

All analyses steps are carried out using the Belle~II analysis software framework~\citep{basf2} and the recorded Belle collision data and Monte Carlo (MC) simulated samples were converted using the tool described in Ref.~\citep{b2b2}. MC samples of \PB meson decays and non-resonant \continuum with $\Pquark=\Pup,\Pdown,\Pstrange,\Pcharm$ continuum processes are generated using the \texttt{EvtGen} generator~\citep{EvtGen}, with sample sizes corresponding to approximately ten and six times the Belle collision data, respectively. The interactions of particles traversing the detector are simulated using \texttt{Geant3}~\citep{Geant3}. QED final state radiation (FSR) is simulated using the \texttt{PHOTOS}~\citep{Photos} package. The \btolng signal is simulated using the calculation of Ref.~\citep{PhysRevD.61.114510}. Charmless semileptonic decays are one of the main contributions to the background in the analysis, in particular \btopln with $\pi^0 \to \gamma \gamma$ can mimic the signal final state. The \btopln background is modeled using the BCL form factor parametrization~\citep{Bourrely:2008za} with central values and uncertainties from the global fit of Ref.~\citep{HFLAV16}.  The remaining charmless semileptonic background is modeled using a mix of resonant and non-resonant modes: the resonant contributions for \mbox{$\PBplus \to \omega \, \ell^+ \, \nu_\ell$}, \mbox{$\PBplus \to \rho^0 \, \ell^+ \, \nu_\ell$}, \mbox{$\PBzero \to \rho^- \, \ell^+ \,  \nu_\ell$} are simulated according to a pole model documented in Ref.~\citep{EvtGen}. The contributions from \mbox{$\PBplus \to \eta \, \ell^+ \, \nu_\ell$}, \mbox{$\PBplus \to \eta' \, \ell^+ \, \nu_\ell$}, \mbox{$\PBplus \to f_2(1270) \, \ell^+ \, \nu_\ell$}, and \mbox{$\PBplus \to b_1(1235) \, \ell^+ \, \nu_\ell$} are modeled using the \texttt{ISGW2} model~\citep{Scora:1995ty}. Non-resonant contributions are modeled using the \texttt{DFN} calculation~\citep{DeFazio:1999ptt} with a choice of its parameters $\lambda_1^{\rm SF}$ and $\bar \Lambda^{\rm SF}$ to approximatively reproduce the first and second moments of the inclusive $m_{X_{\Pup}}$ distribution. The fragmentation of the $X_{\Pup}$ system is performed by \texttt{Jetset}~\citep{SJOSTRAND199474}. Background processes from leptonic, semileptonic or hadronic \PB meson decays are included in the simulation and all relevant decay branching fractions are corrected to correspond to the values of Ref.~\citep{pdg:2017}. 
The efficiencies in the MC are corrected using data-driven methods and are described later. 

The presence of a neutrino in \btolng decays prohibits the full reconstruction of the signal \PB meson, and thus the entire $\PUpsilonFourS$ decay chain is considered to infer the missing momentum of the neutrino. First, a signal electron or muon and a photon candidate with at least $ \SI{1}{GeV} $ in the laboratory frame are identified. The FEI algorithm then hierarchically reconstructs the rest of the event (ROE): The tag-side \PB meson (\btag) is reconstructed using 29 explicit hadronic decay channels, leading to $\mathcal{O}(10000)$ final states. An optimized implementation of gradient-boosted decision trees (BDT) \citep{Keck2017} is used for multivariate classification at the individual stages of the \btag reconstruction, which progresses by forming intermediate particles for ($\PJpsi, \Ppizero, \PKshort, \PD, \PDs, \PDsstar$) from stable particle candidates for ($\APelectron, \APmuon, \PKp, \Pgpp, \gamma$) to reconstruct $\PB$ candidates in six distinct stages. To each reconstructed candidate a signal probability $ P_\mathrm{FEI} $ is assigned, which is calculated by the respective classifier on the properties of the candidate (such as the invariant mass or vertex fit information)  to discriminate signal from background candidates. A detailed description of the entire algorithm can be found in Ref.~\citep{Keck:2018lcd} and references therein. By reconstructing the signal side first, the FEI trains specifically to recognize \btolng decays in conjunction with hadronic \PB meson decays. Using the kinematic information of the \btag, the four-momenta of the signal side \PB meson (\bsig) and thus the missing neutrino can be reconstructed in the center-of-mass frame as 
\begin{align}
 p_{\bsig} = \left( \begin{matrix}  \sqrt{s}/2 \\ - \vec p_{\btag}  \end{matrix} \right) \, \quad \text{and} \quad
 p_{\nu} = \left(  p_{\bsig} - p_\ell - p_\gamma \right) \, , \nn
\end{align}
where $\sqrt{s}$ denotes the center-of-mass energy of the colliding $\APelectron \Pelectron$ pair, and $p_{x}$ and $\vec p_{x}$ are the four- and three-momentum of a given particle $x$. 

Only events with not more than 12 tracks are selected as the signal side consists of only one charged track and signal events typically have a low track multiplicity.
Photons are identified as energy depositions in the calorimeter without an associated track. Only photons with an energy deposition of \mbox{\Eg $ > \SI{100}{MeV}$, $\SI{150}{MeV}$, and $\SI{50}{MeV}$} in the forward end-cap, backward end-cap and barrel part of the calorimeter, respectively, are considered. 
Charged tracks are required to have a distance of closest approach to the nominal interaction point transverse to and along the beam axis of $|\text{d} r| < \SI{2}{cm}$ and $|\text{d} z| < \SI{4}{cm}$, respectively. Charged tracks are identified as electron or muon candidates by combining the information of multiple subdetectors into a likelihood ratio, the lepton identification ($\mathcal{L}_\mathrm{LID}$). For electrons the identifying features are the ratio of the energy deposition in the ECL with respect to the reconstructed track momentum, the energy loss in the CDC, the shower shape in the ECL, the quality of the geometrical matching of the track to the shower position in the ECL, and the photon yield in the ACC~\citep{HANAGAKI2002490}. Muon candidates are identified from charged track trajectories extrapolated to the outer detector. The identifying features are the difference between expected and measured penetration depth as well as the transverse deviation of KLM hits from the extrapolated trajectory~\citep{ABASHIAN200269}. 
Charged tracks are identified as pions or kaons using a likelihood classifier using information from the CDC, ACC, and TOF subdetectors. 

Electrons can radiate sizable fractions of their kinetic energy through bremsstrahlung and FSR processes. Candidates for such photons are identified in the ECL using a cone of 5 degrees around the initial trajectory of the electron candidate, such that only photons radiated near the interaction region can be found. The bremsstrahlung and FSR photons are required to have an energy of \Eg $ < \SI{1}{GeV}$ and if several photon candidates are identified, only the photon with the highest energy is used. The four-momentum of the signal side electron candidate is then corrected by adding the photon energy accordingly.
The prompt signal-photon candidates from the \btolng decay are required to have \EgOneGev in the rest frame of the \bsig meson. Further, signal-photon candidates must provide $R_{9/25} > 0.9$, which is defined as the ratio of the energies deposited in the $3 \times 3$ with respect to the energy deposited in the $5 \times 5$ CsI(Tl) crystals around the maximal energy deposition. $ \mathcal{L}_\mathrm{LID} > 0.8 $ is required for electron and muon candidates. Since the efficiencies of the $\mathcal{L}_\mathrm{LID}$ requirement can differ between MC and data, an efficiency correction is applied, measured on four-lepton and inclusive $ \PB \to X \PJpsi \, (\to \ell^+ \ell^-) $ decays in bins of lepton momentum in the laboratory frame and polar angle. Furthermore, the invariant mass $M_{\PB}$  of the reconstructed lepton-photon pair has to be within $ \SI[parse-numbers=false]{(1.0, 6.0)}{GeV}$. 

Before the FEI tagging algorithm is applied, the corresponding tag-side is cleaned to remove events which do not allow for a reasonable tag-side reconstruction. Photon candidates must provide $R_{9/25} > 0.9$. In addition, a cut on the difference between the beam energy and the energy of the ROE calculated in the center-of-mass frame, $ \Delta E_\mathrm{ROE} < \SI{2.0}{GeV}$, is applied. The beam-constrained mass $ M_\mathrm{bc, ROE} = \sqrt{(\sqrt{s}/2)^2 - \vec{p\,}_\mathrm{\,ROE}^2}$ of the ROE calculated in the center-of-mass frame has to be larger than $ \SI{4.8}{GeV} $. 

After the reconstruction of both \bsig and \btag, the \PUpsilonFourS candidate can be reconstructed. A best-candidate selection based on $ P_\mathrm{FEI} $ (of the \btag candidate) is performed, if more than one candidate per event is reconstructed. Further combinatoric background is removed with cuts on the reconstructed invariant mass of the \PUpsilonFourS candidate of $ M_{\PUpsilonFourS} \in \SI[parse-numbers=false]{(7.5, 10.5)}{GeV}$ (considering the missing neutrino), and the  difference between the beam energy and the energy of the \btag candidate, $\Delta E \in \SI[parse-numbers=false]{(-0.15, 0.1)}{GeV}$. The beam-constrained mass $ M_\mathrm{bc} = \sqrt{(\sqrt{s}/2)^2 - \vec{p\,}_{\,\btag}^2}$ of the \btag candidate has to be within $ \SI[parse-numbers=false]{(5.27,5.29)}{GeV}$. 
Only events with no unassigned tracks (either for \bsig or \btag) and $ E_\mathrm{ECL} \leq \SI{0.9}{GeV}$, where  $ E_\mathrm{ECL}$ is the sum of the remaining unassigned energy depositions in the ECL, are retained. Continuum background is removed by applying an additional cut $ P_\mathrm{FEI} > 0.01 $, whose value was chosen by studying the $ M_\mathrm{bc} $ sideband, defined as  $ M_\mathrm{bc} \in \SI[parse-numbers=false]{(5.24,5.27)}{GeV}$.

The background from \btopln and \btoetaln is peaking in $M_\mathrm{miss}^2$ and is suppressed in a two-step procedure: First, the signal-side photon is combined with any other photon in the ROE to reconstruct a $ \Pgpz $ candidate. The event is vetoed if the candidate satisfies $ M_{\gamma \gamma} \in \SI[parse-numbers=false]{(110, 160)}{MeV}$, where $ M_{\gamma \gamma}$ is the invariant mass of the two photons combined. Further, a multivariate method is trained to suppress the remaining \btopln and the \btoetaln background using the following variables: the number of ECL cluster hits used for the signal-side photon reconstruction, $R_{9/25}$, the lateral distribution of the energy of the ECL cluster hits, the angle between the signal-side photon and the missing momentum $ \vec{p}_{\nu} $ calculated in the rest frame of the \bsig, $ E_\mathrm{ECL} $, and the energy asymmetry, revealing the asymmetry in energy distribution of the lepton and photon candidate of the \bsig, calculated as 
\begin{align}
A =  \frac{E_{\ell} E_{\gamma}}{E_{\ell} + E_{\gamma}}. 
\end{align}

To improve control over the normalization of the peaking background, control samples for \mbox{\btopln}, with $\ell = e,\mu$ are reconstructed. The signal side selection is slightly adapted for the \btopln selection: instead of a single photon with $E_\gamma > 1$ GeV, two photon candidates are combined to form a \Pgpz candidate and only events with an invariant mass of $M_{\gamma \gamma} \in \, (115, 152) \text{MeV}$ (corresponding to approximately $ \pm 3 \sigma $ in $ \pi^0 $ mass resolution), are retained. Both control samples and the \btolng signal decays are analyzed simultaneously to extract the desired signal yields and to constrain the peaking \btopln contaminations in the signal candidates. 

For both the \btolng and the \btopln selections, non-resonant continuum processes are suppressed using a multivariate approach with the aforementioned implementation of BDT. The event topology for continuum processes differs from that of \PB meson decays. This can be exploited to suppress continuum events by using event shape variables, such as the magnitude of the thrust of final state particles forming the \bsig and ROE candidates, the angle between the \bsig and the $z$-axis and between the \bsig and the ROE, the reduced Fox-Wolfram moment $R_2$, the modified Fox-Wolfram moments \citep{SFW} and CLEO Cones \citep{cleocones}. 

The cuts on the multivariate classifier for continuum and the peaking background suppression are simultaneously optimized with Punzi's figure of merit \citep{punzi}. After all selection steps, we obtain a signal reconstruction efficiency for \btolng decays of $ 0.64 \%$ ($ 0.67 \%$) for the electron (muon) final state. On the normalization sample we obtain an efficiency of $ 0.38 \%$ for both final states for \btopln decays. 

To discriminate the signal from background decays, the missing mass squared \Mmisssq of the event is calculated as
\begin{align}\label{eq:mm2}
M_{\rm miss}^2 =  m_{\nu}^2 = p_{\nu}^2  = \left( p_{\bsig} - p_\ell - p_X \right)^2 \, ,
\end{align}
with $p_X $ denoting $p_\gamma$ for \btolng signal events, and $p_{\pi}$ for \btopln normalization events, respectively. The signal and background yields are then obtained using the statistical analysis described in Section~\ref{sec:stats}. The analysis procedure is validated using two signal-depleted sidebands: an off-resonance sample, recorded \SI{40}{MeV} below the $\PUpsilonFourS$ resonance, and the $ M_\mathrm{bc}$ sideband were analyzed. Both showed good agreement between data and the MC expectation.

\section{Statistical Analysis and Limit Setting Procedure}\label{sec:stats}

Signal and background yields are extracted using a binned maximum likelihood fit of the $M_{\rm miss}^2$ distribution. For an individual channel, the likelihood function is constructed as 
\begin{align} \label{eq:likelihood}
 \mathcal{L}_{c} = \prod_i^{\rm bins} \, \mathcal{P}( n_i ; \nu_i )\, ,
\end{align}
with $\mathcal{P}( n_i ; \nu_i )  = \nu_i^{n_i}/\left( n_i!\right) e^{- \nu_i}$ denoting the Poisson distribution with $n_i$ and $\nu_i$ the number of observed and expected events in a given bin $i$ of $M_{\rm miss}^2$, respectively.
Three different likelihood fits are carried out in this manuscript:
\begin{itemize}
 \item[i.] Semileptonic \btoDln decays are analyzed to determine a calibration factor for the FEI tagging efficiency. The selection and obtained calibration factors are further discussed in Section~\ref{sec:fei}. 
 \item[ii.] The branching fraction of \btopln events is determined as a cross check of the FEI calibration procedure, cf. Section~\ref{sec:fei}. 
 \item[iii.]The \btolng signal events are analyzed using a simultaneous fit to the \btolng and \btopln \Mmisssq distributions. A global likelihood function is constructed as
 \begin{align}\label{eq:totlikelihood}
  \mathcal{L}  = \prod_c \mathcal{L}_c \times \prod_k^{\rm syst} \, \mathcal{G}(\theta_k)
 \end{align} 
 with $c$ denoting the reconstructed event type corresponding to the four categories defined by the \btoeng, \btomung,  \btopen and \btopmun channels. Further, $\mathcal{G}(\theta_k)$ denotes the standard normal distribution for nuisance parameters  $\theta_k$, which incorporate systematic uncertainties into the likelihood function. The various systematic uncertainties are further discussed in Section~\ref{sec:syst}.
\end{itemize}
The expected number of events in a given bin $i$ of the $M_{\rm miss}^2$ distribution and in a given category is constructed as
\begin{align}
 \nu_i = \sum_j \nu_j f_{ij} \, ,
\end{align}
with $\nu_j$ the total number of events of type $j$ and $f_{ij}$ denoting the expected fraction of events of type $j$ in the $i^{\rm th}$ bin. The fractions $f_{ij}$ are obtained from the MC simulation and the event types for the \btoDln and \btopln fits are further detailed in Sections~\ref{sec:fei}. For the search for the rare \btolng decay the yield of four event types are used as free parameters in the fit:
\begin{itemize}
 \item[i.] \btolng signal events.
 \item[ii.] \btopln normalization events.
 \item[iii.] \btoxuln background events.
 \item[iv.] Other \PB meson or continuum background events.
\end{itemize}
The \btopln normalization mode is linked between the \btolng and \btopln categories and the global likelihood function $\mathcal{L}$ is maximized to determine the estimates for the number of signal events. Confidence intervals are constructed using the profile likelihood method
\begin{align} \label{eq:proflikelihood}
 \lambda(\nu_j) = \ln \frac{ \mathcal{L}(  \nu_j,  \widehat {\boldsymbol \nu}_{\nu}, \widehat {\boldsymbol \theta}_{\nu})  }{ \mathcal{L}(  \widehat \nu_j, \widehat {\boldsymbol \nu}, \widehat {\boldsymbol \theta})  } \, ,
\end{align}
where $\widehat \nu_j$, $\widehat {\boldsymbol \nu}$ and  $\widehat {\boldsymbol \theta}$ are the values of the normalization of interest, the remaining normalizations, and nuisance parameters that unconditionally maximize the likelihood function while $\widehat {\boldsymbol \nu}_{\nu}$ and $\widehat {\boldsymbol \theta}_{\nu}$ are the values of the other normalizations and nuisance parameters which maximize the likelihood under the condition that the observable of interest is kept fixed at a given value $\nu_j$. In the asymptotic limit, approximate confidence intervals (CI) can be constructed using
\begin{align}\label{eq:CL}
  1 - \mathrm{CI} = \int_{ -2  \lambda(\nu_j) }^{\infty} f_{\chi^2}(x; 1 \, \text{dof})\, \text{d} x \, ,
\end{align}
with $f_{\chi^2}(x; 1 \, \text{dof})$ denoting the $\chi^2$ distribution with one degree of freedom. In case of two parameters of interest a two-dimensional confidence level (CL) can be constructed via Eq.~\ref{eq:CL} by modifying $ \lambda(\nu_j)$ to a likelihood ratio depending on two parameters, $\nu_j$ and $\nu_k$, $ \lambda(\nu_j, \nu_k)$, and with $f_{\chi^2}$ correspondingly then having two degrees of freedom.

In case we observe no significant signal, we set a Bayesian limit by converting the likelihood Eq.~\ref{eq:totlikelihood}, $\mathcal{L} = \mathcal{L}({\boldsymbol n} | \nu_j )$, into a probability density function $\mathcal{F}$ of the parameter of interest $\nu_i$ using a flat prior $\pi(\nu_j)$ such that
\begin{align} \label{eq:bayesian}
 \mathcal{F}(\nu_j | {\boldsymbol n}) = \frac{ \mathcal{L}({\boldsymbol n} | \nu_j ) \pi(\nu_j) }{ \int_{0}^{\infty} \mathcal{L}({\boldsymbol n} | \nu_j ) \pi(\nu_j) \, \text{d} \nu_j  } \, ,
\end{align}
with $\pi(\nu_j) = \text{constant}$ for $\nu_j > 0$ and zero otherwise. In Eq.~\ref{eq:bayesian}, $ {\boldsymbol n}$ denotes the vector of observed event yields in the given bins in all channels. 

The fit procedure was validated using ensembles of pseudoexperiments generated with different input branching fractions for \btolng and \btopln decays. No biases or undercoverage of CI are observed. 

\section{Hadronic tagging efficiency Calibration and \btopln branching fraction}\label{sec:fei}

\begin{figure}[t]
	\begin{center}
			\includegraphics[width=0.5\textwidth]{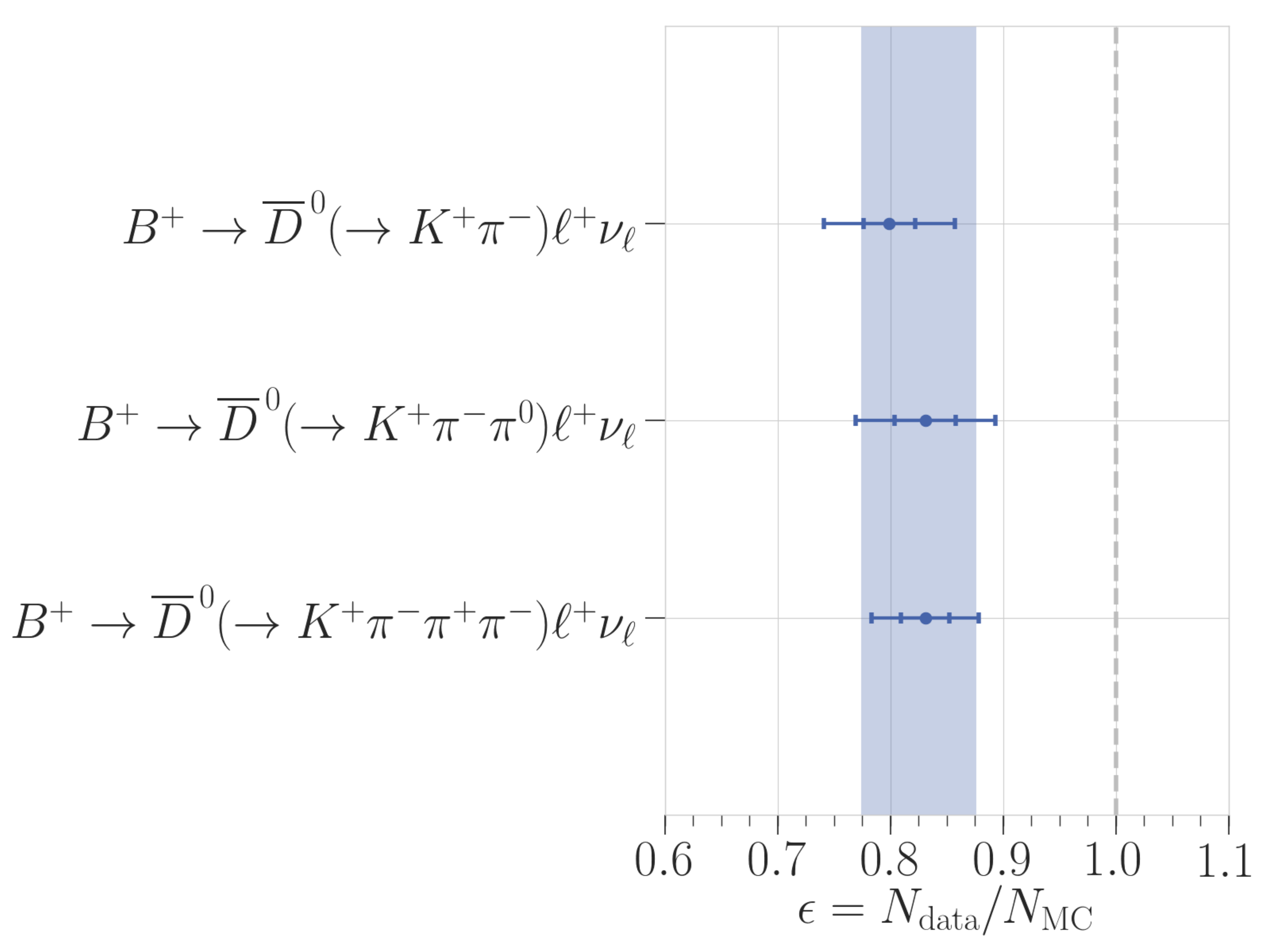}
	\end{center}
	\caption{Extracted calibration factors for the applied tagging algorithm found for the three different semileptonic calibration channels. The inner and outer interval represents the statistical and systematic uncertainty, respectively. The blue band depicts the $ 1 \sigma $ uncertainty band of the averaged calibration factor.}
	\label{fig:calibration}
\end{figure}

The multivariate classifiers that enter the hadronic tagging algorithm of the FEI are trained on simulated events. Due to imperfections in these simulations, e.g. due to inadequate modeling of hadronic decays or experimentally poorly constrained branching fractions, the tagging algorithm exhibits a different performance on simulated and recorded events. Due to this, the tagging algorithm has to be calibrated on data using well known processes. In this analysis, such a calibration is derived using three semileptonic $ \PB $ channels with different multiplicities,
\begin{itemize}
	 \item[i.] $\PBplus \to \APDzero (\to \PKp \Pgpm) \,\ell^{\,+} \Pnulepton$
	 \item[ii.] $\PBplus \to \APDzero (\to \PKp \Pgpm \Pgpz)\,\ell^{\,+} \Pnulepton$
	 \item[iii.] $\PBplus \to \APDzero (\to \PKp \Pgpm \Pgpp \Pgpm)\, \ell^{\,+} \Pnulepton$,
\end{itemize}
where $\ell = e, \mu$. The tag-side selection is identical with the nominal analysis. The signal side selects electron and muon candidates using the criteria detailed in Section~\ref{sec:selection}. Charged pion tracks are required to originate near the IP with $|\text{d} r| < \SI{2}{\cm}$ and $|\text{d} z| < \SI{4}{\cm}$. Charged kaons and pions are separated using a likelihood ratio,  $ \mathcal{P}_{\PK \Pgp} $, which combines the relevant information from the ACC, TOF and CDC subdetectors. We require $ \mathcal{P}_{\PK \Pgp} < 0.4$ for pion and $ \mathcal{P}_{\PK \Pgp} > 0.6$ for kaon candidates, respectively. Neutral pions are reconstructed from the combination of two photons with an invariant mass of $M_{\gamma \gamma} \in \SI[parse-numbers=false]{(117.8, 152)}{MeV}$. $ \APDzero $ candidates are reconstructed combining the kaon candidates with the charged and neutral pion candidates. The reconstructed $ \APDzero $ candidates are required to have $M_{\PD} \in \SI[parse-numbers=false]{(1.858, 1.872)}{GeV}$, $ M_{\PD} \in \SI[parse-numbers=false]{(1.849, 1.879)}{GeV}$ and $ M_{\PD} \in \SI[parse-numbers=false]{(1.854, 1.872)}{GeV}$ for the three channels i. - iii., respectively. Additional loose cuts are applied on the beam-constrained mass of the \bsig candidate (reconstructed from the $ \APDzero $ and the lepton), $ M_\mathrm{bc} > \SI{4.5}{GeV}$, and the cosine of the angle between the true $ \PB $ meson (calculated from beam energy and momentum) and the reconstructed $ \PD \ell $ system, $ |\cos(\theta_{\PB \PD \ell})| < 3.0 $. An unconstrained vertex fit is applied on the $ \PD $ and $ \PB $ candidates and candidates with a $ p $-value of the fit of $ p_{\chi^2} > 0.01 $ are retained.
 
The tag efficiency calibration factor is calculated by extracting the number of signal decays on data and comparing to the expected number of events from the MC simulation. The signal yield is determined using a binned maximum likelihood fit (cf. Section~\ref{sec:stats} ) of the \Mmisssq distribution, reconstructed as 
\begin{align}
M_{\rm miss}^2 =  m_{\nu}^2  = \left( p_{\bsig} - p_\ell - p_{\PD} \right)^2 \,. 
\end{align}
The obtained calibration factors of the three channels are shown in Fig.~\ref{fig:calibration} and the global calibration factor is found to be $ \epsilon = 0.825 \pm 0.014\, \text{(stat.)} \pm 0.049\, \text{(syst.)} $.

To validate the found calibration factor, we measure the branching fraction of the \btopiln decay and compare it to the current world average. We obtain \mbox{$ \mathcal{B}(\btopiln)  = (7.8 \pm 0.6\, (\text{stat.})) \times 10^{-5}$}, which is in agreement with the average $ \mathcal{B_{\mathrm{PDG}}}(\btopiln)  = (7.80 \pm 0.27) \times 10^{-5}$ of Ref.~\citep{pdg:2017}.

\section{Systematic Uncertainties}\label{sec:syst}

There are several systematic uncertainties that affect the measured yields and partial branching fractions: Table~\ref{tab:sys_comb} summarizes the most important sources of uncertainty for the \btolng and \btopln branching fraction measurements.

The effect of all systematic uncertainties are directly incorporated into the likelihood Eq.~\ref{eq:likelihood} via the replacement of
\begin{align}
\nu_j \,  f_{ij} \to \nu_j \,  f_{ij} \times \prod_{k}^{\rm syst} \left( 1 + \theta_k \epsilon_{ijk} \right), 
\end{align}
and
\begin{align}
\nu_j \,  f_{ij} \to \nu_j \,  f_{ij} +  \sum_{k}^{\rm syst} \theta_k \epsilon_{ijk} \, ,
\end{align}
for multiplicative and additive uncertainties, respectively. Nuisance parameters $\theta_k$ are constrained using standard normal distributions $\mathcal{G}(\theta_k)$ in Eq.~\ref{eq:totlikelihood} for relative and absolute uncertainties $\epsilon_{ijk}$ of a source $k$ for a component $j$ and a given bin $i$. Systematic and statistical uncertainties are separated from each other using scans of the likelihood contour in which the systematic nuisance parameters are kept fixed at their best fit value. 

The largest multiplicative systematic uncertainty on both branching fractions stems from the uncertainty on the tagging calibration (see the previous section). It is evaluated by shifting the central value of the combined correction factor according to its statistical and systematic uncertainty. This results in a relative uncertainty of $ 6.2 \% $. The second largest uncertainty for \btopln is given by the statistical uncertainty on the signal reconstruction efficiency. Its uncertainty is evaluated using binomial uncertainties, following the prescription of Ref.~\citep{eff_unc}. Another large multiplicative uncertainty stems from the $\mathcal{L}_\mathrm{LID}$ efficiency, which is corrected in the simulation using data-driven methods. The statistical and systematic uncertainty on these correction factors are propagated and result in an uncertainty of  $ 1.81 \% $ and $ 1.97 \% $ for $ \Delta \mathcal{B} (\btolng)$ and $\mathcal{B} (\btopln) $, respectively. The remaining two multiplicative uncertainties are from the number of $ \PB \APB $ pairs, used to convert the measured yield into (partial) branching fractions, and the uncertainty on reconstruction efficiency differences between the simulation and recorded collisions of charged tracks. The tracking efficiency differences are studied using $ \PDstar \to \PDzero \Ppi $ decays with $ \PDzero \to \Ppi \Ppi \PKs $  and $ \PKs\to \Ppiplus \Ppiminus$. The uncertainty on $N_{ \PB \APB}$ results in a relative error of $1.37 \% $ and for the tracking efficiency an uncertainty of $ 0.35 \% $ for the single signal side track is found.  

\begin{table}[ht!]
	\small
	\centering
	\caption{Systematic uncertainties for the simultaneous fit of both final states.}
	\begin{tabular} {l c c} \toprule
	& \pidecayBR & \mydecayDeltaBR \\ 
	Source& in $ 10^{-5} $ & in $ 10^{-6} $ \\ 
	 \hline
	 Calibration 							& $ \pm 0.49$ & $ \pm 0.09$ \\
	 Reconstruction efficiency & $ \pm 0.20$ & $ \pm 0.01$ \\  
	 $\mathcal{L}_\mathrm{LID}$ efficiency							& $ \pm 0.16$ & $ \pm 0.02$ \\
	 $ N_{\PB \APB} $						& $ \pm 0.11$ & $ \pm 0.02$ \\ 
	 Tracking efficiency					& $ \pm 0.03$ & $ \pm 0.0$ \\
	
	 Peaking background BDT & $ \pm 0.02$ & $ \pm 0.24$ \\
	 PDF templates & $ \pm 0.08$ & $ \pm 0.18$ \\	 
	 BCL model & $ \pm 0.25$ & $ \pm 0.01$\\
	 Reconstructed tag channel & $ \pm 0.01$ & $ \pm 0.14$ \\ 
	 $\PB \to X_{\Pup} \ell^+ \nu_{\ell}$  &$ \pm 0.02$ & $ \pm 0.07$ \\
	 Signal model & $ \pm 0.00$ & $ \pm 0.03$\\
	
	 \textbf{Combined} & $ \mathbf{\pm 0.62} $ & $ \mathbf{\pm 0.36} $ \\ \hline 
	\bottomrule
	\end{tabular}
	\label{tab:sys_comb}
\end{table}

\begin{figure*}[ht!]
	\begin{minipage}{0.49\textwidth} 
		\includegraphics[width=0.98\textwidth]{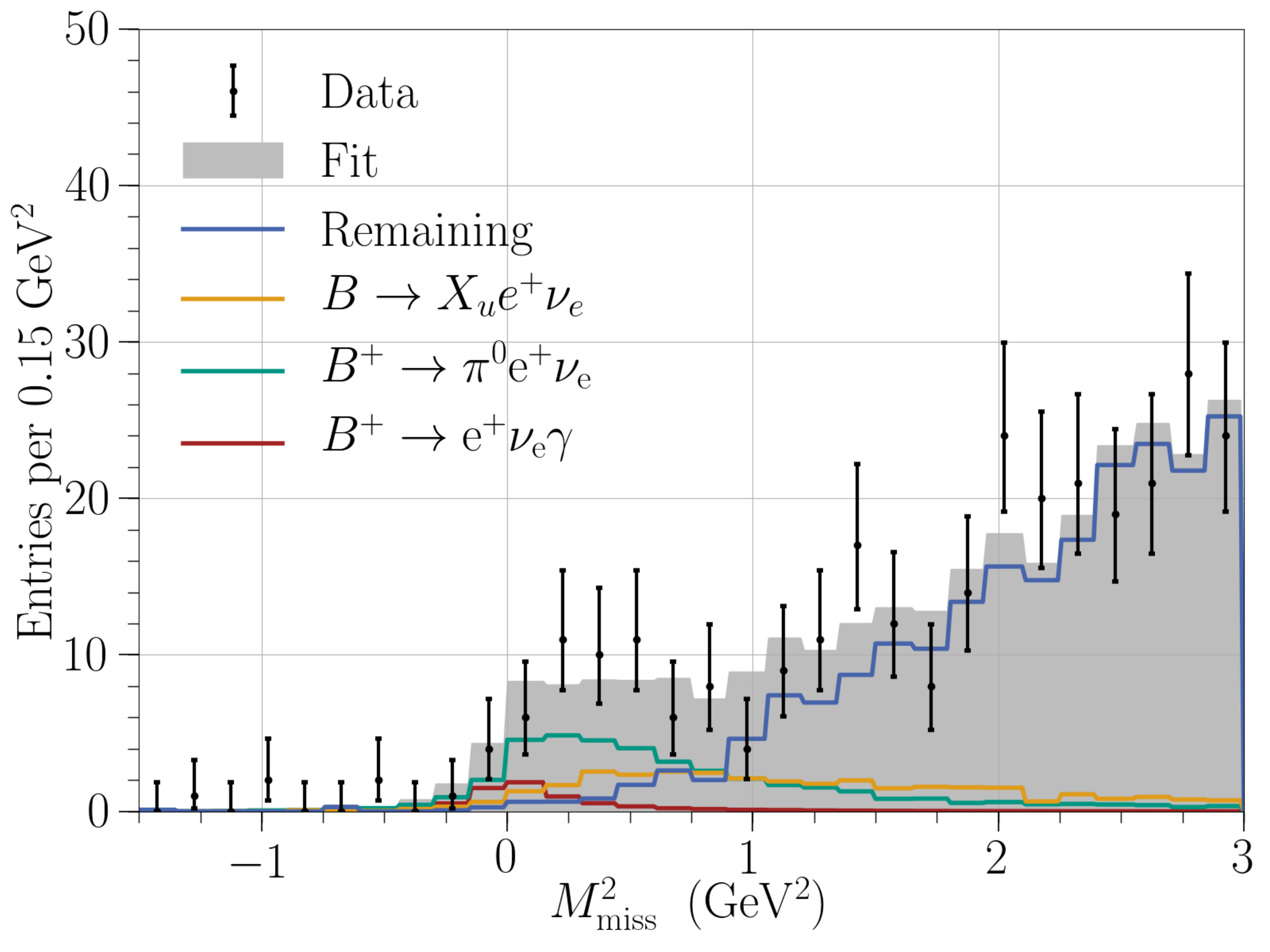}
		\subcaption{\btolngElectron final state}
	\end{minipage}
	\hfill
	\begin{minipage}{0.49\textwidth} 
		\includegraphics[width=0.98\textwidth]{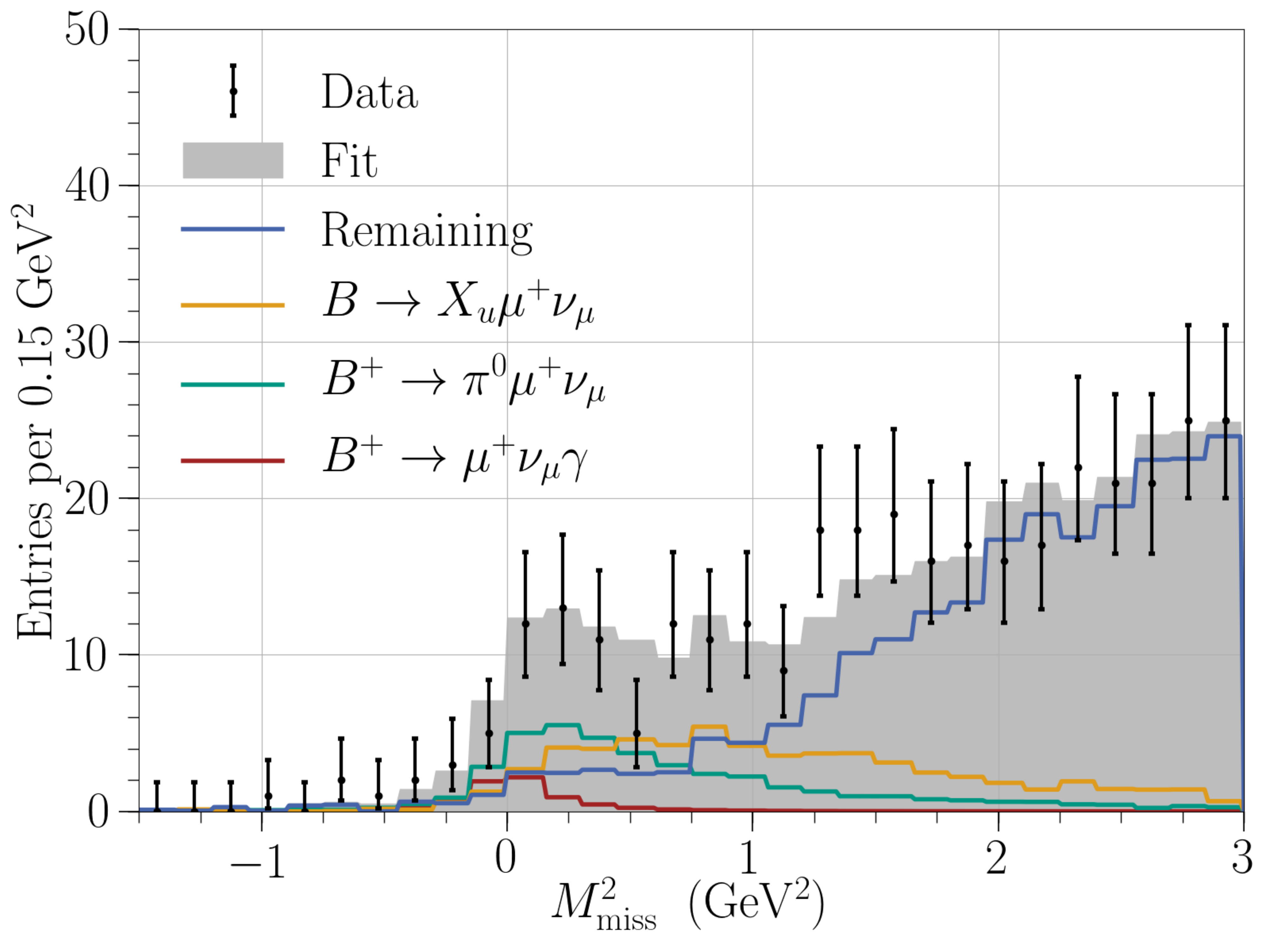}
		\subcaption{\btolngMuon final state}
	\end{minipage}
	\begin{minipage}{0.49\textwidth} 
		\includegraphics[width=0.98\textwidth]{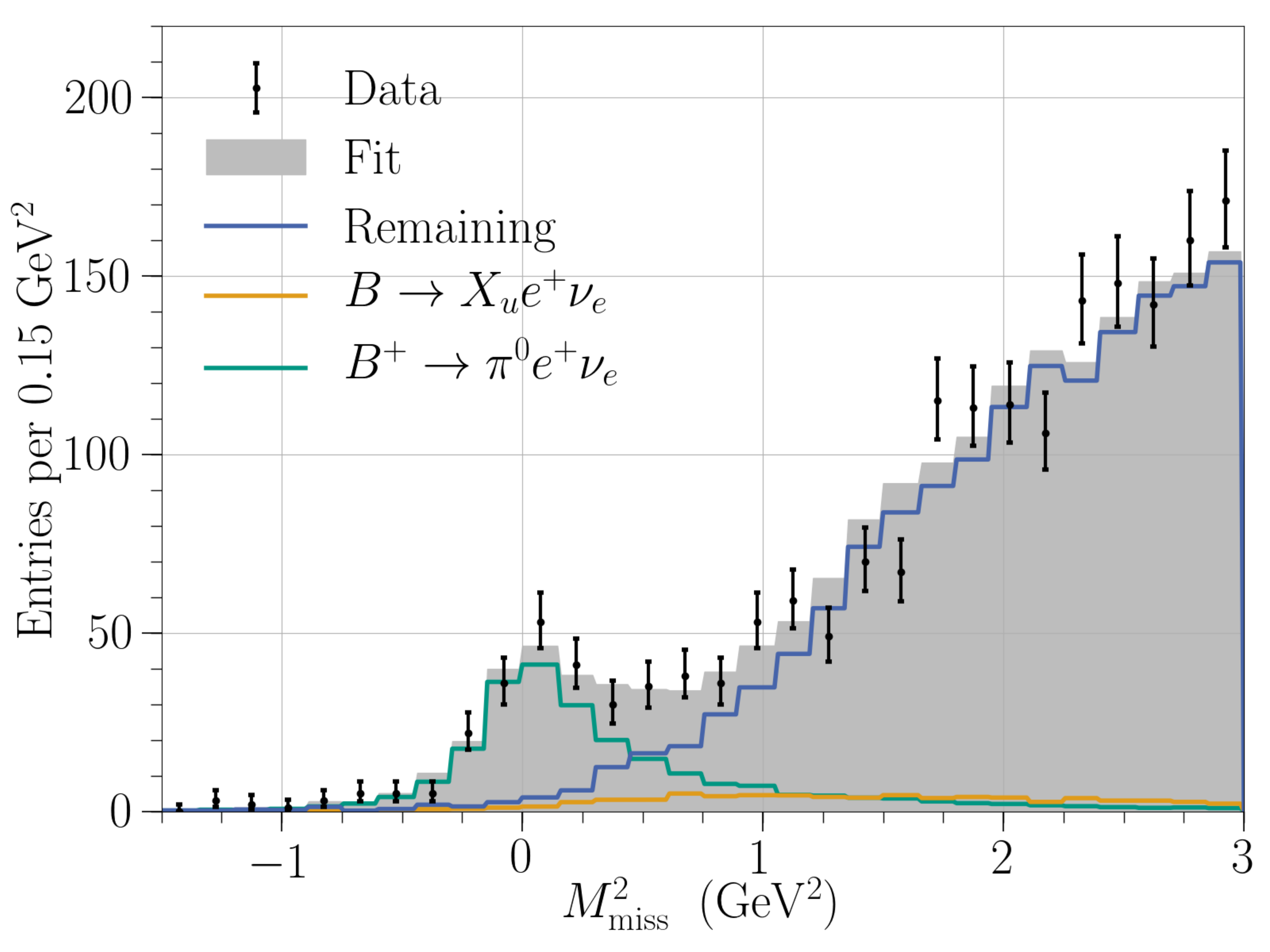}
		\subcaption{\btopen final state}
	\end{minipage}
	\hfill
	\begin{minipage}{0.49\textwidth} 
		\includegraphics[width=0.98\textwidth]{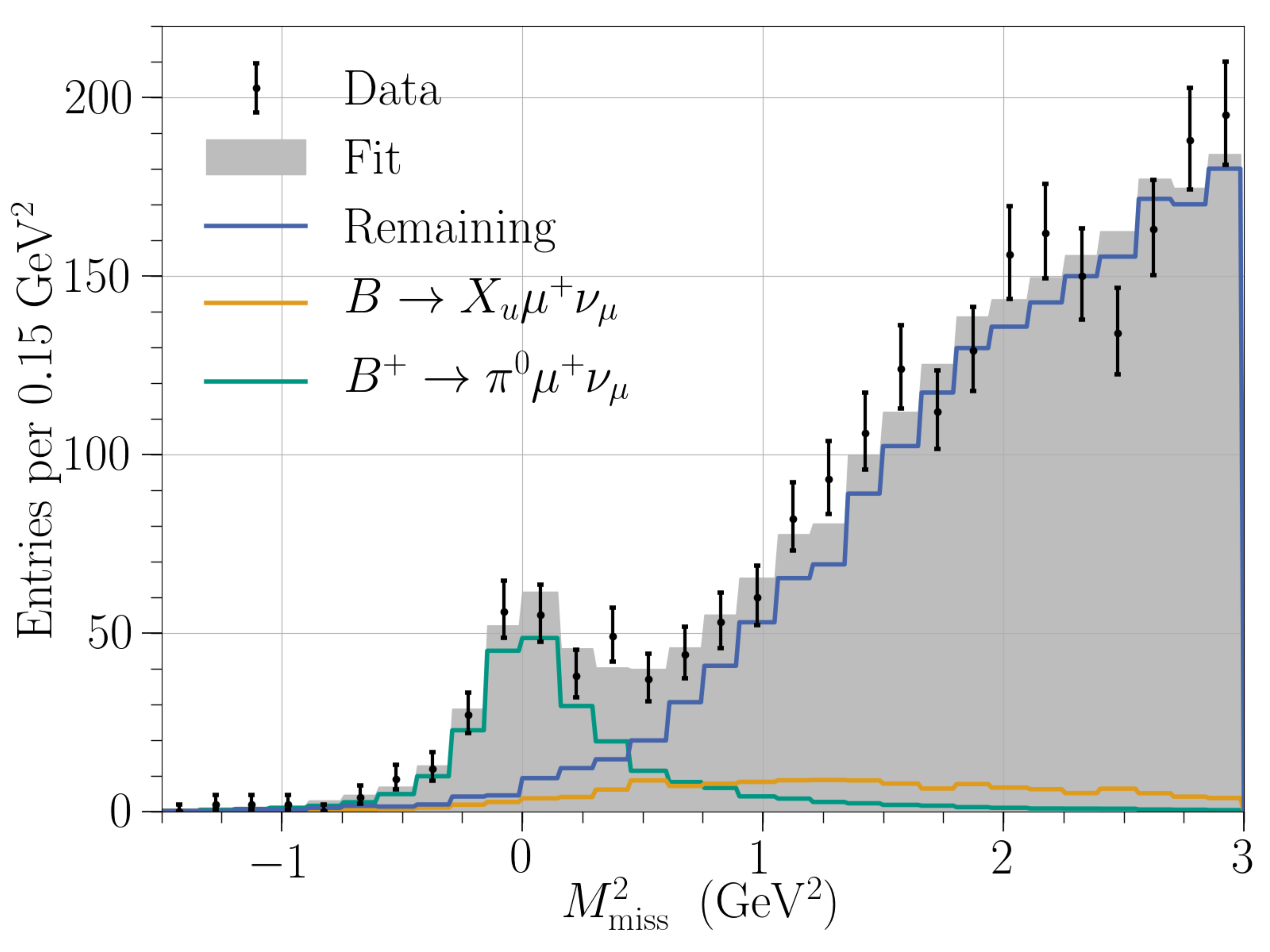}
		\subcaption{\btopmun final state}
		\label{fig:pilnFit}
	\end{minipage}	
	\caption{
	The post-fit \mmsq distributions for the simultaneous fit to the four categories are shown (cf. Section~\ref{sec:stats}). The individual fit components are shown as colored histograms, and the filled gray histogram shows their sum.
	}
	\label{fig:lngammaFit}
\end{figure*}

The largest additive systematic uncertainty for the \btolng partial branching fraction measurement stems from the systematic uncertainty assigned to the multivariate method that suppresses peaking background contributions. This uncertainty is evaluated by reweighting the MC samples to the distribution of the input variables used for the classification on data. The distribution which gives the largest deviation from the nominal result is used to estimate the uncertainty. The second largest additive uncertainty for the \btolng partial branching fraction measurement is due to limited MC statistics. The uncertainty is evaluated for each MC sample individually by producing a large ensemble of templates, where the numbers of entries are varied using a Poisson distribution. The templates of the ensemble are used to repeat the fit to estimate the total uncertainty. The largest additive systematic uncertainty for the \btopln branching fraction is given by the uncertainty on the BCL form factors and is evaluated by variations using the covariance matrix from the global fit of Ref.~\citep{HFLAV16}.

The remaining additive uncertainties on both channels are evaluated as follows: The fraction of the individual channels in which the \btag is reconstructed differs between MC and data. To estimate the impact of this mismatch, the MC samples are corrected to the fraction in data of the reconstructed tag channels and the difference is taken as an estimation for the systematic uncertainty. In the fit, the individual branching fractions of charmless semileptonic background decay modes are kept fixed and modeled as a single floating background template. To estimate uncertainties due to slight shape differences in $M_{\rm miss}^2$ from these templates, we vary the decay branching fractions of \mbox{$\PBplus \to \omega \, \ell^+ \, \nu_\ell$}, \mbox{$\PBplus \to \rho^0 \, \ell^+ \, \nu_\ell$}, \mbox{$\PBzero \to \rho^- \, \ell^+ \,  \nu_\ell$}, \mbox{$\PBplus \to \eta \, \ell^+ \, \nu_\ell$}, \mbox{$\PBplus \to \eta' \, \ell^+ \, \nu_\ell$}, and  \mbox{$\PBzero \to \pi^- \, \ell^+ \,  \nu_\ell$} individually within their uncertainties~\citep{pdg:2017}. The uncertainty on the \btolng signal model is estimated by correcting the simulated events from the prediction of Ref.~\citep{PhysRevD.61.114510} to the state-of-the-art prediction of Ref.~\citep{Beneke:2011nf} and repeating the fit.

 \begin{figure*}[ht!]
	\centering
	\begin{minipage}{0.49\textwidth} 
		\includegraphics[width=0.98\textwidth]{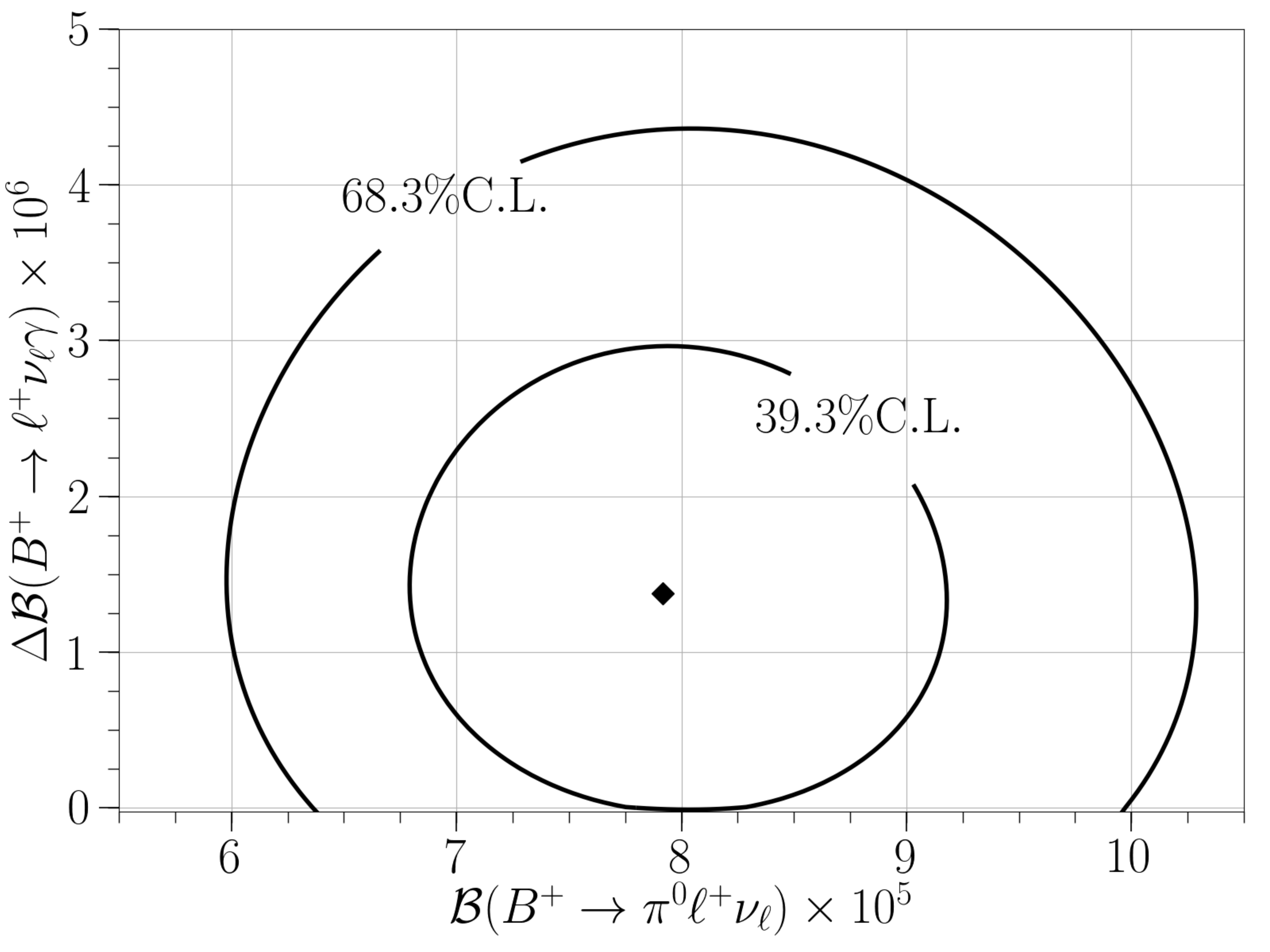}
		\subcaption{Two-dimensional likelihood scan}
		\label{fig:llscansa}
	\end{minipage}	
	\begin{minipage}{0.49\textwidth} 
		\includegraphics[width=0.98\textwidth]{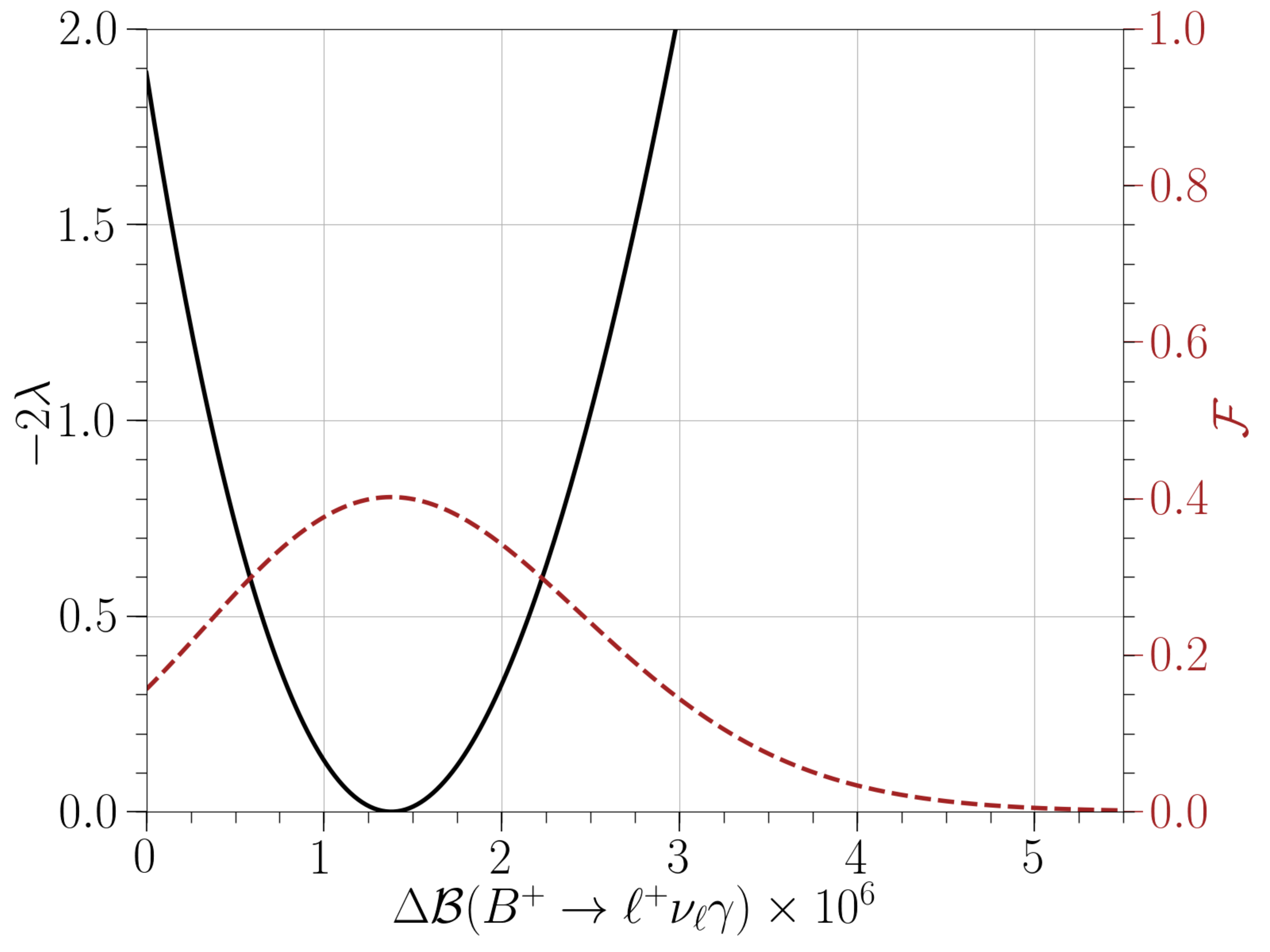}
		\subcaption{One-dimensional likelihood scan and Bayesian PDF}
		\label{fig:llscansb}
	\end{minipage}		
	\caption{Plot (a) shows the two-dimensional likelihood ratio contour $-2 \lambda $ for the combined measurement of \btolng and \btopln. The ellipses correspond to the given confidence level, including systematic uncertainties. Plot (b) shows the one-dimensional likelihood contour and its conversion into a Bayesian PDF $ \mathcal{F}(\nu_j | {\boldsymbol n})$ using a flat prior for the \btolng measurement, see Section~\ref{sec:stats} for details. }
	
\end{figure*}

\section{Results}\label{sec:results}

Figure~\ref{fig:lngammaFit} shows the \mmsq distribution of the selected data events in the four categories of \btolngElectron, \btolngMuon, \btopen, and \btopmun. The selected events are used to maximize the likelihood function Eq.~\ref{eq:likelihood} numerically, determining the four (\btolng) and three (\btopln) event types detailed in Section~\ref{sec:stats}.

The fitted \btolng signal, \btopln normalization and other background contributions are shown as colored histograms and the summed signal plus background template is shown as a filled gray histogram. The observed partial branching fraction of \btolng with $E_{\gamma} > 1$ GeV is 
\begin{align}
 \Delta \mathcal{B}(\btolng)  & = \left( 1.4 \pm 1.0 \pm 0.4 \right) \times 10^{-6}  \, , 
\end{align}
where the first error is statistical and the second error contains all systematic uncertainties discussed in Section~\ref{sec:syst}. The significance over the background-only hypothesis for the \btolng signal, as calculated using the likelihood ratio, is 1.4 standard deviations. The \btopln branching fraction is found to be
\begin{align}
  \mathcal{B}(\btopln)  & = \left( 7.9 \pm 0.6 \pm 0.6 \right) \times 10^{-5} \, ,
 \end{align} 
 and has better statistical precision than the measurement of Ref.~\citep{Sibidanov:2013rkk}\footnote{The statistical overlap with the previous measurement is unknown. Since the current result is not measured in bins of $ q^2 $, the previous result should still be used for the determination of \VubAbs and world averages of the branching fraction.}. A summary of all fit results, including fits of the individual electron and muon samples, is presented in Table~\ref{tab:fitres}. Figure~\ref{fig:llscansa} shows the two-dimensional likelihood ratio contours of $-2 \lambda $ (see Eq.~\ref{eq:proflikelihood}) for both branching fractions. The correlation between  $\Delta \mathcal{B}(\btolng)$ and $\mathcal{B}(\btopln)$ is found to be $\rho = -2.7\%$.
 
Due to the low significance of the measured \btolng signal, we convert the likelihood into a Bayesian probability density function (PDF), with the procedure detailed in Section~\ref{sec:stats}. Figure~\ref{fig:llscansb} shows the one-dimensional likelihood ratio scan and the obtained Bayesian PDF, which was obtained using a flat prior in the partial branching fraction. The resulting limit for \btolng at 90\% CL is
 \begin{align}
 \Delta \mathcal{B}(\btolng)  < 3.0 \times 10^{-6} \, \text{at} \,\, 90\% \, \text{CL} \, .
 \end{align}
 This provides a significantly more stringent limit than previous searches, and a summary of previous limits and individual limits for the electron and muon signal channel can be found in Table~\ref{tab:limits}.
 
\begin{figure}[h!]
	\centering
	\includegraphics[width=0.51\textwidth]{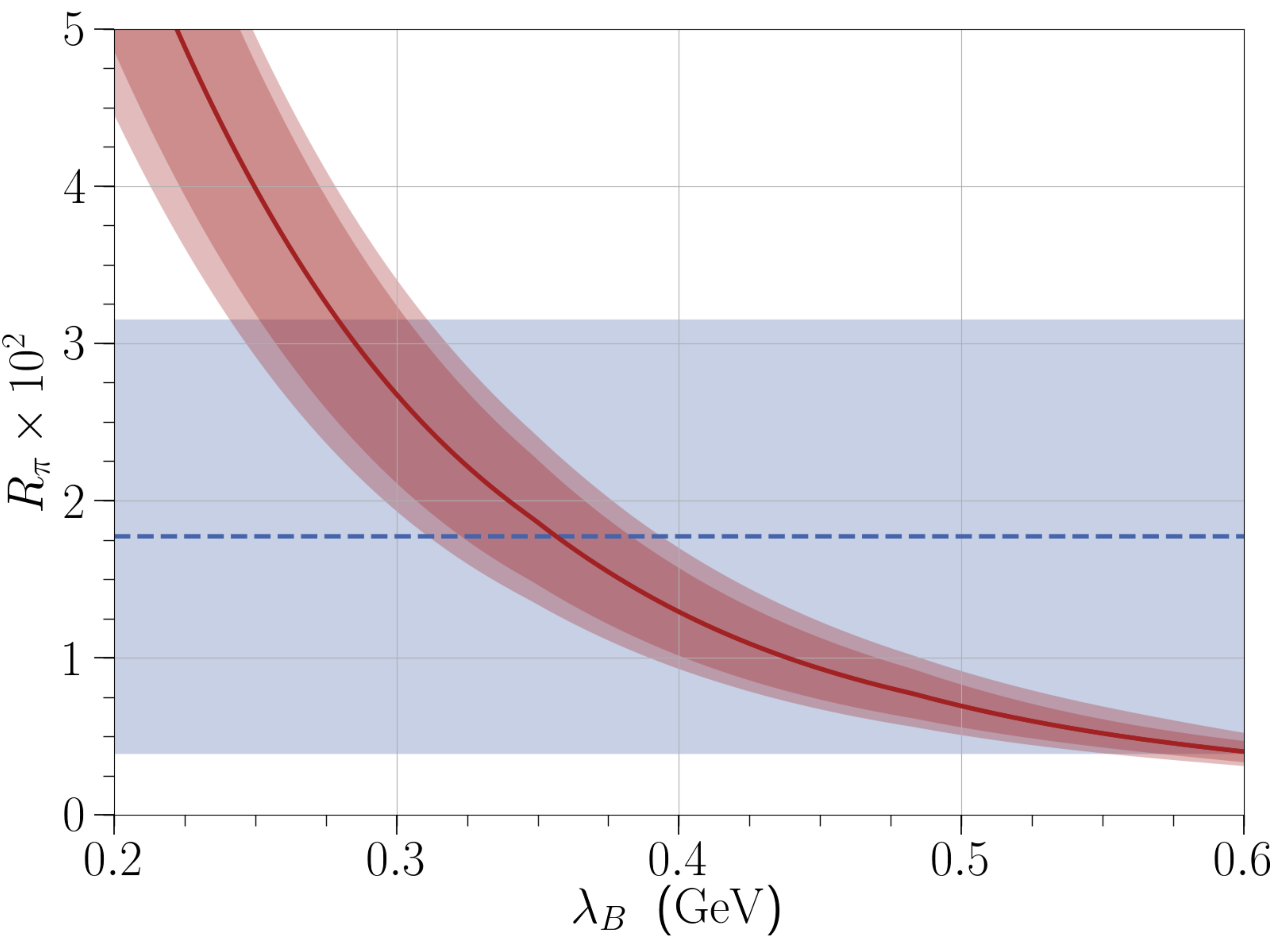}
	\caption{ The theory prediction of Refs.~\citep{Beneke:2018wjp} and \citep{HFLAV16} (red line with  $1\sigma$ uncertainties) for $R_{\pi}$ is compared to the measured value and $1\sigma$ uncertainty (blue dashed line and band). The dark red band shows the theoretical uncertainty, the light red band additionally contains the light-cone distribution amplitude model dependence.}
	\label{fig:rlam}
\end{figure}

Using the \btolng and \btopln branching fractions, the first inverse moment $\lb$ of the leading-twist \PB meson light-cone distribution amplitude $\phi_+$ can be determined. Instead of directly using the measured \btolng partial branching fraction, we use the theoretically well understood \btopln decay rate to derive a measurement of \lb which is independent of \Vub. The value of \lb is related to this ratio as
\begin{align} \label{eq:rpi}
 R_{\pi} = \frac{ \Delta \mathcal{B}(\btolng)}{  \mathcal{B}(\btopln)} = \frac{ \Delta \Gamma(\lb) }{ \Gamma(\btopln) } \, ,
\end{align}
with $\Delta \Gamma(\lb)$ denoting the partial decay rate as a function of \lb with \Eg $> 1$ GeV, and $\Gamma(\btopln)$ denoting the total decay rate of \btopln. Using the central values and the full experimental covariance we measure
\begin{align}
 R_{\pi}^{\rm meas} = \left( 1.7 \pm 1.4 \right) \times 10^{-2} \, .
\end{align}
For the prediction of the \btopln decay rate, we use the global fit \citep{HFLAV16} of \mbox{$\Gamma(\btopln) = \left| V_{ub} \right|^2 \times \left( 2.4 \pm  0.2 \right) \times 10^{-12}$ GeV}. For the partial   
\btolng decay rate the predictions and uncertainties of Ref.~\citep{Beneke:2018wjp} extrapolated to \Eg $> 1$ GeV are used. In Ref.~\citep{Beneke:2018wjp} three different models are used to evaluate the dependence of the partial decay rate on the functional form of the light-cone distribution amplitude. Figure~\ref{fig:rlam} shows the predicted and measured $R_\pi$ ratio as a function of \lb. We solve Eq.~\ref{eq:rpi} numerically and in Table~\ref{tab:lb} the determined value of \lb for each of the three models are given, including the corresponding theoretical uncertainties of Ref.~\citep{Beneke:2018wjp}. We use the shift in the central value between all three models to also quote a value of \lb, whose uncertainty should incorporate the overall model dependence. For this we find
\begin{align}
 \lb = 0.36 \substack{+0.25 \\ -0.08}\substack{+0.03 \\ -0.03}\substack{+0.03 \\ -0.03} \, \text{GeV} = 0.36 \substack{+0.25 \\ -0.09} \, \text{GeV} \, ,
\end{align}
where the first uncertainty is experimental, the second from the theoretical uncertainty on the \btolng prediction of Ref.~\citep{Beneke:2018wjp} and the \btopln uncertainty from Ref.~\citep{HFLAV16}, and the third uncertainty is due to the light-cone distribution amplitude model dependence. We further obtain a one-sided limit of
\begin{align}
 \lb > 0.24 \, \text{GeV} \, 
\end{align}
at 90\% CL. 

Note, that these estimates might suffer from additional uncertainties from the extrapolation to \Eg $> 1$ GeV. Further details can be found in Ref.~\citep{Beneke:2018wjp}.

\begin{table}[t]
\footnotesize
\centering
\caption{Measured central values and the corresponding significance for the different channels.}
\begin{tabular} {c |c c | c c } \toprule
%
$\ell$ & 
$ \mathcal{B}$(\pidecay) $ (10^{-5}) $ & $ \sigma $ & $ \Delta \mathcal{B}$(\mydecay) $(10^{-6})$ & $ \sigma $  
 
 \\   \hline
 $ \Pe$ 			& $8.3\substack{+0.9 \\ -0.8} \pm 0.9$ & 8.0 	   & $1.7\substack{+1.6 \\ -1.4} \pm 0.7$ & 1.1 \\
 $ \Pgm$  	        & $7.5\substack{+0.8 \\ -0.8} \pm 0.6$ & 9.6	   & $1.0\substack{+1.4 \\ -1.0} \pm 0.4$ & 0.8 \\
 $ \Pe,\Pgm$        & $7.9\substack{+0.6 \\ -0.6} \pm 0.6$ & 12.6 	   & $1.4\substack{+1.0 \\ -1.0} \pm 0.4$ & 1.4 \\
 \label{tab:fitres}
\end{tabular}
\end{table}

\begin{table}[t]
	\small
	\centering
	\caption{Comparison to previous results of the partial branching fraction measurement. All limits correspond to the $ 90\% $ CL.}
	\label{tab:compare_res}
	\begin{tabular} {c |  c  c  c}
		\toprule
		\multirow{2}{*}{} & \multicolumn{3}{c}{$\Delta \mathcal{B}$(\mydecay) limit $(10^{-6})$} \\
		\cmidrule(l{0.5em}){2-4} 
		$\ell$ 		& BaBar \citep{PhysRevD.80.111105} & Belle \citep{Heller:2015vvm} & This work \\ \hline			
		$ \Pe $		& - 	  & $< 6.1 $ & $< 4.3 $ \\
		$ \Pgm $	& - 	  & $< 3.4 $ & $< 3.4 $ \\
		$ \Pe,\Pgm$	& $< 14 $ & $< 3.5 $ & $< 3.0 $ \\
		\bottomrule
	\end{tabular}
	\label{tab:limits}
\end{table}

\begin{table}[t]
\small
\centering
\caption{The determined values of \lb using the predictions of Ref.~\citep{Beneke:2018wjp} are given. A detailed description of the three approaches to model the functional form of the light-cone distribution amplitude (LCDA) can be found in Ref.~\citep{Beneke:2018wjp}. The first uncertainty are experimental and the second from theory. }
\begin{tabular} {c  | c c c  } \toprule
 & \lb (GeV) \\ \hline 
 Model I   &  $0.36 \substack{+0.25 \\ -0.08} \substack{+0.03 \\ -0.03}$ \\   
 Model II  &  $0.38 \substack{+0.25 \\ -0.06} \substack{+0.05 \\ -0.08}$ \\   
 Model III &  $0.32 \substack{+0.24 \\ -0.07} \substack{+0.05 \\ -0.08}$ \\
\end{tabular}
  \label{tab:lb}
\end{table}

\section{Summary}\label{sec:summary}

In this manuscript, an improved search for the radiative leptonic decay \btolng on the full Belle data set recorded at the \PUpsilonFourS resonance is presented. The results improve the previous analysis by our collaboration and increase the signal efficiency by a factor of three. In addition, the description of the important \btopln background was improved, by analyzing simultaneously \btopln signal events and using the global fit result of Ref.~\citep{HFLAV16} to describe its form factors. The large improvement in sensitivity stems from employing a newly developed tagging algorithm developed for the Belle~II experiment, the Full Event Interpretation~\citep{Keck:2018lcd}.  Although this drastically improves the sensitivity, no significant signal of \btolng decays is observed. As it is not possible to determine the statistical overlap with the previous Belle result, this work supersedes Ref.~\citep{Heller:2015vvm}.

The determined partial branching fraction for \btolng decays with photon energies $E_\gamma > \SI{1}{GeV}$ in the \bsig rest frame is found to be
\begin{align}
 \Delta \mathcal{B}(\btolng)  & = \left( 1.4 \pm 1.0 \pm 0.4 \right) \times 10^{-6}  \, ,
\end{align}
with a significance of 1.4 standard deviations over the background-only hypothesis. Using the likelihood contour and a flat prior, we determine a Bayesian upper limit of
 \begin{align}
 \Delta \mathcal{B}(\btolng)  < 3.0 \times 10^{-6}  \, ,
 \end{align}
at 90\% confidence level.

In addition, we report an improved determination of the first inverse momentum \lb of the the light-cone distribution amplitude of the $ \PB $ meson. It is done using a \Vub independent way, by normalizing the measured partial branching fraction to the branching fraction of \btopln. This reduces the experimental uncertainties, and the theoretical prediction of the total decay width of \btopln is well understood. 

Using the result of Ref.~\citep{Beneke:2018wjp}, its associated uncertainties and an additional uncertainty to assess the model dependence, we obtain
\begin{align}
 \lb = 0.36 \substack{+0.25 \\ -0.09} \, \text{GeV} \, 
\end{align}
or $\lb > 0.24 \, \text{GeV}$ at 90\% CL. 

The search of \btolng is limited by the available data set and its sensitivity will be greatly enhanced by the upcoming Belle~II experiment. The anticipated data set of 50 ab${}^{-1}$ will greatly reduce the experimental uncertainties on \lb. Together with recent developments in lattice QCD calculations reducing the theoretical uncertainties on \lb, the feasibility of \btolng increases as an alternative channel to measure \VubAbs to provide a consistency check of the SM.

\section{Acknowledgments}

\input{ack.tex}



\end{document}

%% file: shorts.tex
\newcommand{\btag}{\ensuremath{\PB_{\rm{tag}}}\xspace}
\newcommand{\bsig}{\ensuremath{\PB_{\rm{sig}}}\xspace}
\newcommand{\Eg}{\ensuremath{E_{\gamma}}\xspace}
\newcommand{\EgOneGev}{\ensuremath{E_{\gamma}> \SI{1}{GeV}}\xspace}
\newcommand{\lb}{\ensuremath{\lambda_{\PB}}\xspace}
\newcommand{\nn}{\nonumber}
\newcommand{\Mmisssq}{$M_{\rm{miss}}^2$\xspace }

\newcommand{\Vub}{\ensuremath{V_{\Pup \Pbottom}}\xspace}
\newcommand{\VubAbs}{\ensuremath{|V_{\Pup \Pbottom}|}\xspace}

\newcommand{\btolng}{\ensuremath{\PBplus \to \ell^{\,+} \Pnulepton \Pgamma}\xspace}
\newcommand{\btopln}{\ensuremath{\PBplus \to \Pgpz \ell^{\,+} \Pnulepton}\xspace}
\newcommand{\btoetaln}{\ensuremath{\PBplus \to \Peta \ell^{\,+} \Pnulepton}\xspace}
\newcommand{\btolngElectron}{\ensuremath{\PBplus \to \APelectron \Pnue \Pgamma}\xspace}
\newcommand{\btolngMuon}{\ensuremath{\PBplus \to \APmuon \Pnum \Pgamma}\xspace}
\newcommand{\btopen}{\ensuremath{\PBplus \to \Pgpz \APelectron \Pnue}\xspace}
\newcommand{\btopmun}{\ensuremath{\PBplus \to \Pgpz \APmuon \Pnum}\xspace}
\newcommand{\continuum}{\ensuremath{\APelectron \Pelectron \to \Pquark \APquark}\xspace}

\newcommand{\btoDln}{\ensuremath{\PBplus \to \APDzero \ell^+ \nu_{\ell} }\xspace}

\newcommand{\btoeng}{\ensuremath{\PBplus \to \APelectron \Pnue \gamma}\xspace}
\newcommand{\btomung}{\ensuremath{\PBplus \to \APmuon \Pnum \gamma}\xspace}
\newcommand{\btopiln}{\ensuremath{\PBplus \to \Pgpz \Pleptonplus \Pnulepton}\xspace}

\newcommand{\btoxuln}{\ensuremath{\PB \to X_{\Pup} \Pleptonplus \Pnulepton}\xspace}

\newcommand{\BFuplep}{\ensuremath{ \Delta \mathcal{B}( \PB^{+} \to \Plepton^{\, +} \Pnulepton \Pgamma) < 3.0 \times 10^{-6} }\xspace}

\newcommand{\lambdab}{\ensuremath{ \lambda_{\PB} > \SI{0.24}{GeV} }\xspace}

%% file: authors.tex
\noaffiliation
\affiliation{Institut f\"ur Experimentelle Teilchenphysik, Karlsruher Institut f\"ur Technologie, 76131 Karlsruhe}
\affiliation{University of the Basque Country UPV/EHU, 48080 Bilbao}
\affiliation{Beihang University, Beijing 100191}
\affiliation{Brookhaven National Laboratory, Upton, New York 11973}
\affiliation{Budker Institute of Nuclear Physics SB RAS, Novosibirsk 630090}
\affiliation{Faculty of Mathematics and Physics, Charles University, 121 16 Prague}
\affiliation{Chonnam National University, Kwangju 660-701}
\affiliation{University of Cincinnati, Cincinnati, Ohio 45221}
\affiliation{Deutsches Elektronen--Synchrotron, 22607 Hamburg}
\affiliation{University of Florida, Gainesville, Florida 32611}
\affiliation{Key Laboratory of Nuclear Physics and Ion-beam Application (MOE) and Institute of Modern Physics, Fudan University, Shanghai 200443}
\affiliation{Justus-Liebig-Universit\"at Gie\ss{}en, 35392 Gie\ss{}en}
\affiliation{Gifu University, Gifu 501-1193}
\affiliation{II. Physikalisches Institut, Georg-August-Universit\"at G\"ottingen, 37073 G\"ottingen}
\affiliation{SOKENDAI (The Graduate University for Advanced Studies), Hayama 240-0193}
\affiliation{Gyeongsang National University, Chinju 660-701}
\affiliation{Hanyang University, Seoul 133-791}
\affiliation{University of Hawaii, Honolulu, Hawaii 96822}
\affiliation{High Energy Accelerator Research Organization (KEK), Tsukuba 305-0801}
\affiliation{J-PARC Branch, KEK Theory Center, High Energy Accelerator Research Organization (KEK), Tsukuba 305-0801}
\affiliation{Forschungszentrum J\"{u}lich, 52425 J\"{u}lich}
\affiliation{IKERBASQUE, Basque Foundation for Science, 48013 Bilbao}
\affiliation{Indian Institute of Technology Bhubaneswar, Satya Nagar 751007}
\affiliation{Indian Institute of Technology Guwahati, Assam 781039}
\affiliation{Indian Institute of Technology Hyderabad, Telangana 502285}
\affiliation{Indian Institute of Technology Madras, Chennai 600036}
\affiliation{Indiana University, Bloomington, Indiana 47408}
\affiliation{Institute of High Energy Physics, Chinese Academy of Sciences, Beijing 100049}
\affiliation{Institute of High Energy Physics, Vienna 1050}
\affiliation{INFN - Sezione di Napoli, 80126 Napoli}
\affiliation{INFN - Sezione di Torino, 10125 Torino}
\affiliation{Advanced Science Research Center, Japan Atomic Energy Agency, Naka 319-1195}
\affiliation{J. Stefan Institute, 1000 Ljubljana}
\affiliation{Kennesaw State University, Kennesaw, Georgia 30144}
\affiliation{King Abdulaziz City for Science and Technology, Riyadh 11442}
\affiliation{Department of Physics, Faculty of Science, King Abdulaziz University, Jeddah 21589}
\affiliation{Kitasato University, Sagamihara 252-0373}
\affiliation{Korea Institute of Science and Technology Information, Daejeon 305-806}
\affiliation{Korea University, Seoul 136-713}
\affiliation{Kyoto University, Kyoto 606-8502}
\affiliation{Kyungpook National University, Daegu 702-701}
\affiliation{LAL, Univ. Paris-Sud, CNRS/IN2P3, Universit\'{e} Paris-Saclay, Orsay}
\affiliation{\'Ecole Polytechnique F\'ed\'erale de Lausanne (EPFL), Lausanne 1015}
\affiliation{P.N. Lebedev Physical Institute of the Russian Academy of Sciences, Moscow 119991}
\affiliation{Faculty of Mathematics and Physics, University of Ljubljana, 1000 Ljubljana}
\affiliation{Ludwig Maximilians University, 80539 Munich}
\affiliation{Luther College, Decorah, Iowa 52101}
\affiliation{University of Malaya, 50603 Kuala Lumpur}
\affiliation{University of Maribor, 2000 Maribor}
\affiliation{Max-Planck-Institut f\"ur Physik, 80805 M\"unchen}
\affiliation{School of Physics, University of Melbourne, Victoria 3010}
\affiliation{University of Mississippi, University, Mississippi 38677}
\affiliation{University of Miyazaki, Miyazaki 889-2192}
\affiliation{Moscow Physical Engineering Institute, Moscow 115409}
\affiliation{Moscow Institute of Physics and Technology, Moscow Region 141700}
\affiliation{Graduate School of Science, Nagoya University, Nagoya 464-8602}
\affiliation{Universit\`{a} di Napoli Federico II, 80055 Napoli}
\affiliation{Nara Women's University, Nara 630-8506}
\affiliation{National Central University, Chung-li 32054}
\affiliation{National United University, Miao Li 36003}
\affiliation{Department of Physics, National Taiwan University, Taipei 10617}
\affiliation{H. Niewodniczanski Institute of Nuclear Physics, Krakow 31-342}
\affiliation{Niigata University, Niigata 950-2181}
\affiliation{University of Nova Gorica, 5000 Nova Gorica}
\affiliation{Novosibirsk State University, Novosibirsk 630090}
\affiliation{Osaka City University, Osaka 558-8585}
\affiliation{Pacific Northwest National Laboratory, Richland, Washington 99352}
\affiliation{Panjab University, Chandigarh 160014}
\affiliation{Peking University, Beijing 100871}
\affiliation{University of Pittsburgh, Pittsburgh, Pennsylvania 15260}
\affiliation{Theoretical Research Division, Nishina Center, RIKEN, Saitama 351-0198}
\affiliation{University of Science and Technology of China, Hefei 230026}
\affiliation{Seoul National University, Seoul 151-742}
\affiliation{Showa Pharmaceutical University, Tokyo 194-8543}
\affiliation{Soongsil University, Seoul 156-743}
\affiliation{University of South Carolina, Columbia, South Carolina 29208}
\affiliation{Stefan Meyer Institute for Subatomic Physics, Vienna 1090}
\affiliation{Sungkyunkwan University, Suwon 440-746}
\affiliation{School of Physics, University of Sydney, New South Wales 2006}
\affiliation{Department of Physics, Faculty of Science, University of Tabuk, Tabuk 71451}
\affiliation{Tata Institute of Fundamental Research, Mumbai 400005}
\affiliation{Department of Physics, Technische Universit\"at M\"unchen, 85748 Garching}
\affiliation{Toho University, Funabashi 274-8510}
\affiliation{Department of Physics, Tohoku University, Sendai 980-8578}
\affiliation{Earthquake Research Institute, University of Tokyo, Tokyo 113-0032}
\affiliation{Department of Physics, University of Tokyo, Tokyo 113-0033}
\affiliation{Tokyo Institute of Technology, Tokyo 152-8550}
\affiliation{Tokyo Metropolitan University, Tokyo 192-0397}
\affiliation{Virginia Polytechnic Institute and State University, Blacksburg, Virginia 24061}
\affiliation{Wayne State University, Detroit, Michigan 48202}
\affiliation{Yamagata University, Yamagata 990-8560}
\affiliation{Yonsei University, Seoul 120-749}
  \author{M.~Gelb}\affiliation{Institut f\"ur Experimentelle Teilchenphysik, Karlsruher Institut f\"ur Technologie, 76131 Karlsruhe} 
  \author{F.~U.~Bernlochner}\affiliation{Institut f\"ur Experimentelle Teilchenphysik, Karlsruher Institut f\"ur Technologie, 76131 Karlsruhe} 
  \author{P.~Goldenzweig}\affiliation{Institut f\"ur Experimentelle Teilchenphysik, Karlsruher Institut f\"ur Technologie, 76131 Karlsruhe} 
  \author{F.~Metzner}\affiliation{Institut f\"ur Experimentelle Teilchenphysik, Karlsruher Institut f\"ur Technologie, 76131 Karlsruhe} 
  \author{I.~Adachi}\affiliation{High Energy Accelerator Research Organization (KEK), Tsukuba 305-0801}\affiliation{SOKENDAI (The Graduate University for Advanced Studies), Hayama 240-0193} 
  \author{H.~Aihara}\affiliation{Department of Physics, University of Tokyo, Tokyo 113-0033} 
  \author{S.~Al~Said}\affiliation{Department of Physics, Faculty of Science, University of Tabuk, Tabuk 71451}\affiliation{Department of Physics, Faculty of Science, King Abdulaziz University, Jeddah 21589} 
  \author{D.~M.~Asner}\affiliation{Brookhaven National Laboratory, Upton, New York 11973} 
  \author{H.~Atmacan}\affiliation{University of South Carolina, Columbia, South Carolina 29208} 
  \author{T.~Aushev}\affiliation{Moscow Institute of Physics and Technology, Moscow Region 141700} 
% ...
\collaboration{The Belle Collaboration}

%% file: ack.tex
%
%

We thank the KEKB group for the excellent operation of the
accelerator; the KEK cryogenics group for the efficient
operation of the solenoid; and the KEK computer group, and the Pacific Northwest National
Laboratory (PNNL) Environmental Molecular Sciences Laboratory (EMSL)
computing group for strong computing support; and the National
Institute of Informatics, and Science Information NETwork 5 (SINET5) for
valuable network support.  We acknowledge support from
the Ministry of Education, Culture, Sports, Science, and
Technology (MEXT) of Japan, the Japan Society for the 
Promotion of Science (JSPS), and the Tau-Lepton Physics 
Research Center of Nagoya University; 
the Australian Research Council including grants
DP180102629, 
DP170102389, 
DP170102204, 
DP150103061, 
FT130100303; 
Austrian Science Fund under Grant No.~P 26794-N20;
the National Natural Science Foundation of China under Contracts
No.~11435013,  
No.~11475187,  
No.~11521505,  
No.~11575017,  
No.~11675166,  
No.~11705209;  
Key Research Program of Frontier Sciences, Chinese Academy of Sciences (CAS), Grant No.~QYZDJ-SSW-SLH011; 
the  CAS Center for Excellence in Particle Physics (CCEPP); 
the Shanghai Pujiang Program under Grant No.~18PJ1401000;  
the Ministry of Education, Youth and Sports of the Czech
Republic under Contract No.~LTT17020;
the Carl Zeiss Foundation, the Deutsche Forschungsgemeinschaft, the
Excellence Cluster Universe, and the VolkswagenStiftung;
the Department of Science and Technology of India; 
the Istituto Nazionale di Fisica Nucleare of Italy; 
National Research Foundation (NRF) of Korea Grants
No.~2015H1A2A1033649, No.~2016R1D1A1B01010135, No.~2016K1A3A7A09005
603, No.~2016R1D1A1B02012900, No.~2018R1A2B3003 643,
No.~2018R1A6A1A06024970, No.~2018R1D1 A1B07047294; Radiation Science Research Institute, Foreign Large-size Research Facility Application Supporting project, the Global Science Experimental Data Hub Center of the Korea Institute of Science and Technology Information and KREONET/GLORIAD;
the Polish Ministry of Science and Higher Education and 
the National Science Center;
the Grant of the Russian Federation Government, Agreement No.~14.W03.31.0026; 
the Slovenian Research Agency;
Ikerbasque, Basque Foundation for Science, Basque Government (No.~IT956-16) and
Ministry of Economy and Competitiveness (MINECO) (Juan de la Cierva), Spain;
the Swiss National Science Foundation; 
the Ministry of Education and the Ministry of Science and Technology of Taiwan;
and the United States Department of Energy and the National Science Foundation.

We thank the KEKB group for excellent operation of the
accelerator; the KEK cryogenics group for efficient solenoid
operations; and the KEK computer group, the NII, and 
PNNL/EMSL for valuable computing and SINET5 network support.  
We acknowledge support from MEXT, JSPS and Nagoya's TLPRC (Japan);
ARC (Australia); FWF (Austria); NSFC and CCEPP (China); 
MSMT (Czechia); CZF, DFG, EXC153, and VS (Germany);
DST (India); INFN (Italy); 
MOE, MSIP, NRF, RSRI, FLRFAS project and GSDC of KISTI and KREONET/GLORIAD (Korea);
MNiSW and NCN (Poland); MES (Russia); ARRS (Slovenia);
IKERBASQUE and MINECO (Spain); 
SNSF (Switzerland); MOE and MOST (Taiwan); and DOE and NSF (USA).


%% file: Blnugamma.bbl
\begin{thebibliography}{99}
\bibitem{CC}
Throughout this paper, 
the inclusion of the charge-conjugate decay mode is implied unless otherwise stated.
\bibitem{NU}
Natural units are used throughout this paper with $ c = 1 $.
\end{thebibliography}
